\documentclass[journal=jpccck,manuscript=article]{achemso}

\usepackage[version=3]{mhchem}

\usepackage{xr}
\makeatletter
\newcommand*{\addFileDependency}[1]{
  \typeout{(#1)}
  \@addtofilelist{#1}
  \IfFileExists{#1}{}{\typeout{No file #1.}}
}
\makeatother

\newcommand*{\myexternaldocument}[1]{
    \externaldocument{#1}
    \addFileDependency{#1.tex}
    \addFileDependency{#1.aux}
}
\myexternaldocument{SI}

\usepackage{gensymb}
\let\oldAA\AA
\renewcommand{\AA}{\text{\normalfont\oldAA}}

\usepackage[section]{placeins}
\makeatletter
\let\ftype@table\ftype@figure
\makeatother
\usepackage{chemmacros}
\usepackage[section]{placeins}

\author{Amy S. McKeown-Green}
\affiliation{Department of Chemistry, University of California, Berkeley, California
94720, United States}
\author{Justin C. Ondry}
\affiliation{Department of Chemistry, University of California, Berkeley, California
94720, United States}
\alsoaffiliation{Kavli Energy NanoScience Institute, Berkeley, California 94720, United
States}
\author{Michelle F. Crook}
\affiliation{Department of Chemistry, University of California, Berkeley, California
94720, United States}
\alsoaffiliation{Materials Sciences Division, Lawrence Berkeley National Laboratory,
Berkeley, California 94720, United States}
\author{Jason J. Calvin}
\affiliation{Department of Chemistry, University of California, Berkeley, California
94720, United States}
\alsoaffiliation{Materials Sciences Division, Lawrence Berkeley National Laboratory,
Berkeley, California 94720, United States}
\author{A. Paul Alivisatos}
\affiliation{Department of Chemistry, University of California, Berkeley, California
94720, United States}
\alsoaffiliation{Kavli Energy NanoScience Institute, Berkeley, California 94720, United
States}
\alsoaffiliation{Materials Sciences Division, Lawrence Berkeley National Laboratory,
Berkeley, California 94720, United States}
\alsoaffiliation{Department of Materials Science and Engineering, University of California,
Berkeley, California 94720, United States}
\email{amy_mckeown-green@berkeley.edu, jondry@berkeley.edu}

\title{Examining the Role of Chloride Ligands on Defect Removal in Imperfectly Attached Semiconductor Nanocrystals for 1D and 2D Attachment Cases}

\abbreviations{}


\begin{document}

\begin{tocentry}
\includegraphics[scale=.85]{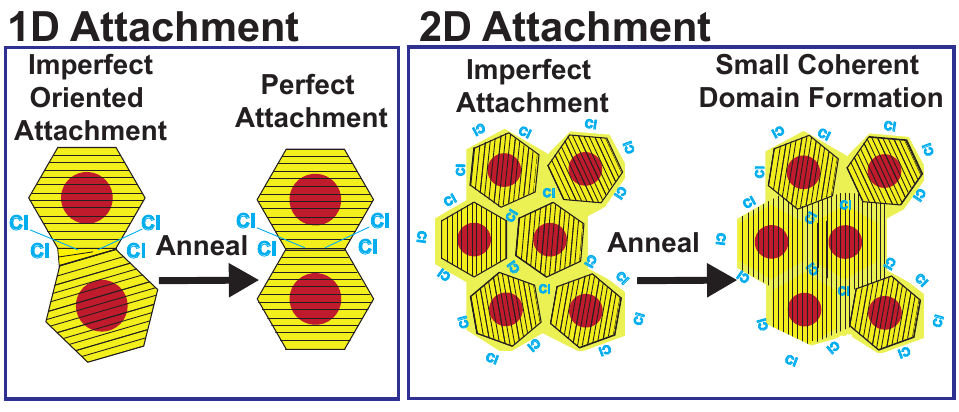}

\end{tocentry}

\begin{abstract}

Semiconducting, core--shell nanocrystals (NCs) are promising building blocks for the construction of higher dimensional artificial nanostructures using oriented attachment. However, the assembly and epitaxial attachment steps critical to this construction introduce disorder and defects which inhibit the observation of desirable emergent electronic phenomena. Consequently, understanding defect formation and remediation in these systems as a function of dimensionality is a crucial step to perfecting their synthesis. In this work, we use \textit{in situ} high resolution transmission electron microscopy to examine the role of chloride ligands as remediator agents for imperfect attachment interfaces between CdSe/CdS core--shell NCs for both 1D and 2D attachment cases. In the 1D case, we find that the presence of chloride additives in imperfectly attached NC dimers can result in defect removal speeds nearly twice as large as those found in their plain, non-chloride treated counterparts. However, when we increased the dimensionality of the system and examined 2D NC arrays, we found no statistically significant difference in attachment interface quality between the chloride and non-chloride treated samples. We propose that this discongruity arises from fundamental differences between 1D and 2D NC attachment and discuss synthetic guidelines to inform future nanomaterial superlattice design. 

\end{abstract}

\section{Introduction} \label{intro}

The diverse size-dependent properties of colloidal nanocrystals (NCs) make them exciting building blocks for the construction of hierarchical nanomaterials with emergent electronic, optical, and mechanical properties.\cite{Xu2013, Yoreo2015, Domenech2019} One interesting class of these hierarchical nanostructures is that of colloidal inorganic NCs which can self-assemble into 2D and 3D super--lattices.\cite{Murray1995,dong_binary_2010,Xu2013} While structurally interesting, the inorganic NC cores in these assemblies are typically separated by long organic ligands, preventing strong electronic coupling. One method to bring the inorganic cores closer together and enable stronger electronic coupling is oriented attachment \textemdash{} the collective alignment and fusion of NCs along their crystallographic facets\cite{Cui2019, B.V.Salzmann2021}. This is desirable as the strong electronic overlap enabled by oriented attachment is predicted to yield exotic optoelectronic properties, such as tunable Dirac cones.\cite{Peng2012, Kalesaki2014,Ondry2021} When epitaxially connected in 2D arrays, core--shell NCs are particularly promising as their core\textendash to\textendash core spacing can be synthetically tuned, thus modifying the degree of electronic coupling.\cite{Ondry2021} Unfortunately, while organic ligand--capped NC assembly is typically a reversible process, oriented attachment is often irreversible due to the strength of inorganic bonds.\cite{Cho2005,Cui2019} As a consequence of this irreversibility, crystal defects (dislocations, stacking faults, \textit{etc.}) can become trapped at the interfaces between NCs,\cite{Penn1998,OndryAccount2021} leading to the formation of electronic trap states which inhibit electronic coupling.\cite{Whitham2016,Ondry2019,Ondry2021} Developing strategies to remove defects in these materials is therefore a crucial step in the bottom--up synthesis of these materials and ultimately, the observation of and control over their unconventional materials properties. 

It has been well-established that the imperfect attachment of PbE and CdE (E = S, Se, Te) NCs often leads to the formation of 1D line defects (dislocations) which can create undesirable mid-gap electronic states.\cite{Ondry2018,Ondry2019, OndryAccount2021} Thermally annealing the materials has shown limited success in removing these undesirable defects.\cite{Ondry2019} This is partly due to a critical caveat of thermal defect removal in attached NCs: success requires that defects are mobile at temperatures below the internal recrystallization temperature of the individual NCs. If this condition is not met, the superperiodicity of the NC superstructure is lost. This limitation of pure thermal annealinng has motivated a search for lower temperature  strategies for remediating NC attachment interfaces and removing defects. 

Chemical treatments are an intriguing possibility for facilitating the removal of structural defects at temperatures lower than those required by thermal annealing. For example, it is well-established that chloride ions cause drastic structural transformation and electronic passivation in II-VI materials upon thermal annealing.\cite{Major2016,Jain2019,M.Norman2014,Krishna2004} More specifically, chloride ions have been observed to increase the rate of grain growth and decrease the temperature of the wurtzite-zinc blende phase transformation in CdE (E = Se,Te) NC thin-films.\cite{M.Norman2014,Zhang2016,Crisp2014} This is relevant as the wurtzite/zinc blende phase transition in II-V materials is thought to proceed \textit{via} the collective movement of partial dislocations,\cite{Zheng2013} suggesting that chloride ligands may act to increase dislocation mobility thereby altering the phase transition kinetics. Overall, the ability of chloride ions to facilitate drastic atomic rearrangement at modest temperatures indicates that they may  be promising agents for facilitating the removal of interfacial defects, such as those which arise from imperfect oriented attachment. 

A final, unexplored factor impacting defect removal is attachment dimensionality (\textit{i.e.,} 1D dimer, 2D sheet, and 3D supercrystal attachment). Thus far, most detailed mechanistic studies on dislocation removal have focused on simple dimer pairs.\cite{Ondry2018, Ondry2019} However, in the case of epitaxially fused superlattices, the number of attachment interfaces of a NC equals the number of nearest neighbors and scales directly with the dimensionality of the system. Consequently, the probability that a single NC will participate in an imperfect attachment case increases as we move from the 1D dimer, to the 2D NC Sheet, and finally the 3D supercrystal attachment case. The current understanding of attachment beyond dimers suggests that geometric frustration further impedes defect removal from imperfect interfaces.\cite{Grason2016,Tanjeem2021} However, the degree to which these defects are irreconcilable remains unknown, and strategies to improve removal kinetics have not been investigated. 

In this work, we explore chloride additives as facilitators of defective interface repair for epitaxially connected, wurtzite CdSe/CdS NCs in both 1D and 2D attachment cases. For the 1D case, we employ \textit{in situ} high resolution transmission electron microscopy (HRTEM) to obtain defect removal speeds and trajectories in imperfectly attached dimers for both the chloride treated and non--chloride treated samples. While the defect type and removal trajectories are identical in both samples, we find the chloride treated sample exhibits higher maximum defect removal speeds. However, there are multiple instances where defect removal is slow or incomplete regardless of chemical treatment. To address this, we propose that a defect's local environment (i.e. degree of interparticle mistilt, substrate interactions, etc...) is a dominating factor in determining whether a dislocation will be successfully removed. In the case that a dislocation's local environment permits removal, chloride additives increase defect mobility and removal rate. For the 2D case, we prepare ordered 2D arrays of wurtzite CdSe/CdS NCs attached via layer--by--layer growth (SILAR)\cite{Ondry2021} which we characterize with HRTEM and selected area electron diffraction (SAED). The angle mistilt between the NCs’ $\{1\bar{1}00\}$ atomic lattice planes is used as a metric to quantify attachment interface quality en masse and is measured using both SAED and a sliding window Fourier transform.  SAED reveals that both samples have pseudo-single crystal grains larger than 50$\mu$m$^2$  with a 15.1$\degree$ $\pm$ 0.6 FWHM of the in--plane, $\{1\bar{1}00\}$ angle distribution irrespective of the presence of chloride additives. At the particle-to-particle limit, we find both chloride and non--chloride treated samples have similar coherent crystallographic domain sizes and distributions which we take as a sign of negligible interface remediation.  We posit this lack of interfacial remediation in the 2D NC arrays despite the presence of chloride additives stems from critical differences in 1D and 2D attachment dynamics. We suggest guidelines for preparing 2D NC arrays with structures that minimize interparticle mistilt and have thermodynamically accessible defect removal pathways. 

\section{Results}

In both 1D and 2D attachment cases, we use well--faceted NCs terminated by $\{1\overline{1}00\}$ planes as their epitaxial attachment is known to result in two distinct cases: a perfect epitaxial interface and an impefect interface containing an edge dislocation with a $b=\frac{a}{3}\langle2\overline{1}\overline{1}0\rangle$ Burgers vector (a = in--plane lattice vector).\cite{Ondry2019} This synthetic control over NC faceting and attachment allows for the facile generation of identical defect-types in both chloride treated and non--chloride treated samples, enabling the role of chloride ligands in dislocation dynamics to be isolated and studied. Importantly, the synthesis of the CdSe/CdS core--shell NCs did not involve any precursors which could be sources of chlorine,\cite{Ondry2021} enabling chloride ligands to be introduced post-synthesis. 

\subsection{Exploring the role of chloride additives on defect removal from imperfect attached CdSe/CdS dimer interfaces} \label{Dimers}

\begin{figure}[htb]
\centering
\includegraphics[scale=0.6]{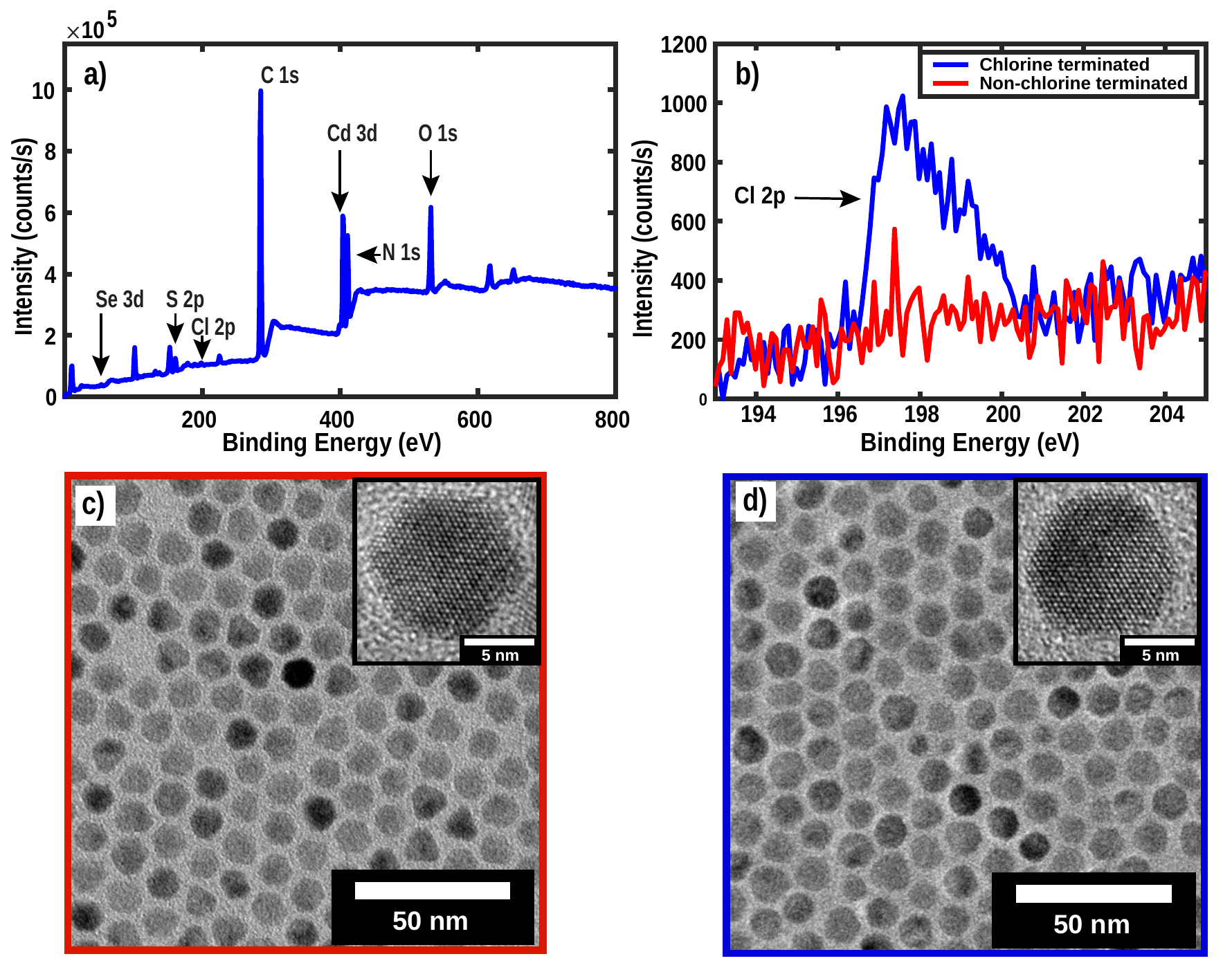}
\caption{ X-ray photoelectron spectra showing the presence of chloride ligands and HRTEM images showing consistent size and shape distributions between both samples. a) X-ray photoelectron survey spectra of the CTAC treated CdSe/CdS core-shell NC sample. b) High-resolution X-ray photoelectron spectra of the Cl 2p peak for both the chloride treated (blue) and non-chloride treated (red) samples showing the presence of chloride in the chloride-treated sample. c) HRTEM image of CdSe/CdS NCs with their native ligands. d) HRTEM image of CdSe/CdS NCs following CTAC treatment and the introduction of chloride ligands.}
\label{fig:XPS}
\end{figure}

In the dimer pairs, cetyltrimethylammonium chloride (CTAC) was used as the chloride source (See Methods).\cite{Saruyama2010} The successful introduction of chloride ligands to the desired sample was confirmed using X-ray photoelectron spectroscopy (XPS). \ref{fig:XPS}a shows a survey XPS spectrum containing NC system component peaks Se 3d, S 2p, Cd 3d, and Cl 2p as well as contaminant C 1s, O 1s, and N 1s peaks. The presence of Cl in the survey is consistent with Cl surface ligands having been successfully incorporated via ligand exchange. \ref{fig:XPS}b shows a high-resolution scan of the Cl 2p peak for both the non-chloride treated sample (red) and the chloride treated sample (blue) confirming the absence of Cl surface ligands prior to the sample’s treatment with CTAC and the successful introduction of Cl ligands via ligand exchange. XPS quantification shows a 0.34\% atomic percent chlorine. In light of the accuracy of such low-concentration calculations, this number should be taken as a representative of the order of magnitude of the concentration. This order of magnitude is likely sufficient to impact the system’s defect dynamics when compared to previous experiments investigating the role of chloride ligands in NC thin--film transformations.\cite{M.Norman2014} \ref{fig:XPS}c,d show HRTEM images of the CdSe/CdS core--shell NCs before (red) and after the CTAC treatment (blue) showing the NCs’ size and shape are not significantly altered by the chloride treatment 

\begin{figure}[htb]
\centering
\includegraphics[scale=1]{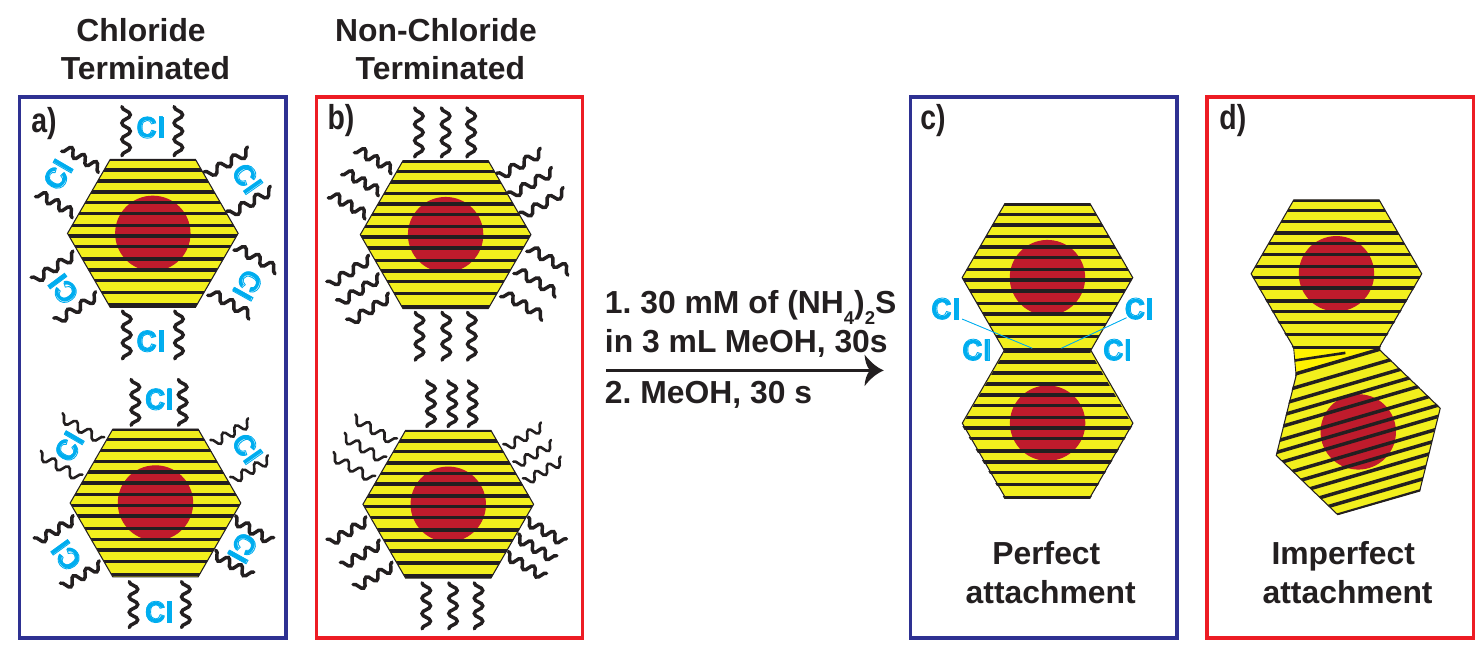}
\caption{Interfacial attachment scheme showing the formation of both a perfect and imperfect interface in dimer CdSe/CdS core--shell NCs. a) CdSe/CdS core--shell NCs treated with CTAC. b) CdSe/CdS NCs with their native ligands. c) Perfect attachment between two NCs with chloride ions trapped at the interface. d) Imperfect attachment between two NCs in the absence of chloride ions. CdSe/CdS core--shell NC dimer attachment is induced by exposure to ammonium sulfide both in the presence of chloride ligands and without. This attachment procedure yielded both perfect and imperfect attachment interfaces in both cases.}
\label{fig:Attachment Scheme}
\end{figure}

NC attachment was induced by dipping the NC--covered TEM grids in a dilute, methanol solution of ammonium sulfide\cite{Zhang2011} resulting in dimers with both perfect and imperfect attachment interfaces. \ref{fig:Attachment Scheme}a and b provide a diagram of the separate chloride treated and non-chloride treated NC samples before attachment. \ref{fig:Attachment Scheme}c shows a scheme in which chloride surface ligands are incorporated in the interface during attachment, possibly enabling a perfect attachment interface to be obtained following remediation. \ref{fig:Attachment Scheme}d shows the possible case where imperfect attachment yields an edge dislocation (an extra half plane of atoms) trapped at the attachment interface of the non-chloride treated sample. 

\textit{In situ} HRTEM was used to record image series of edge dislocation removal dynamics in both chloride treated and non-chloride treated samples for a fixed electron fluence of 8500 $e^-\AA^{-2}s^{-1}$. Controlling the does rate ensured similar effective temperature for all observed trajectories.\cite{Dyck2015,Woehl2019,Kisielowski2015} Imperfectly attached NC dimers were imaged until the defect was removed or the sample was destroyed by the electron beam.\cite{Egerton2004} The lattice resolution of the HRTEM movies allowed the position of the dislocation to be tracked as a function of time to obtain dislocation removal trajectories (see Methods).

\begin{figure}[htb]
\centering
\includegraphics[scale=1]{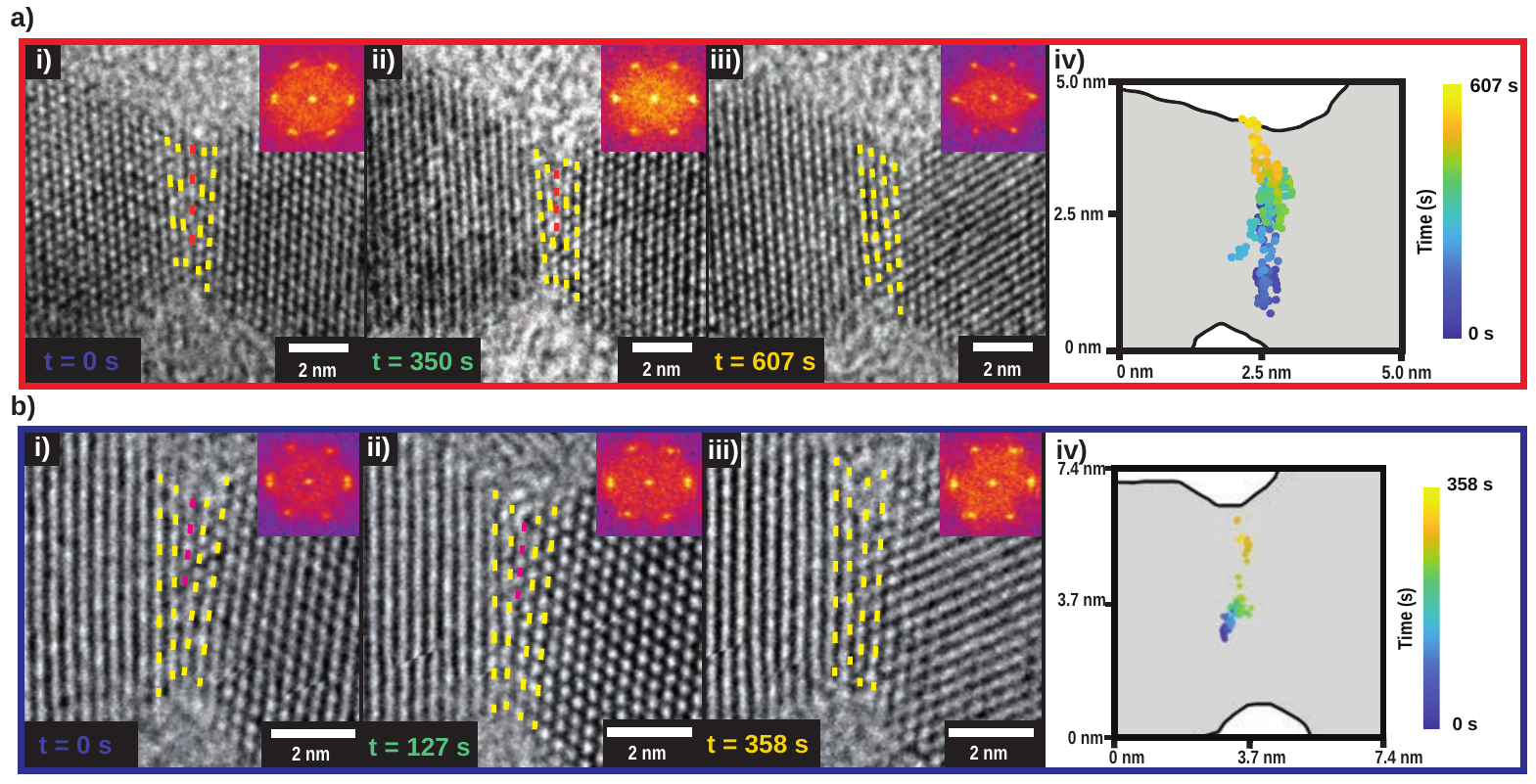}
\caption{Dislocation removal trajectories during electron beam annealing for both the non-chloride and chloride treated samples. i)-iii) snapshots of \textit{in situ} HRTEM time series with inset Fourier transforms and iv) plot of the dislocation core position as a function of time for a) a $b=\frac{a}{3}\langle2\overline{1}\overline{1}0\rangle$ Burgers vector dislocation in a non-chloride treated sample and b) a $b=\frac{a}{3}\langle2\overline{1}\overline{1}0\rangle$ Burgers vector dislocation in a chloride treated sample.}
\label{fig:Trajectories}
\end{figure}

The study was limited to edge dislocations with $b=\frac{a}{3}\langle2\overline{1}\overline{1}0\rangle$ Burgers vectors by design as these are the only defect expected for $\{1\overline{1}00\}$ attachment.\cite{Ondry2019}  \ref{fig:Trajectories}a$/$b i)-iii) provide snapshots from \textit{in situ} HRTEM movies showing dislocation removal in the non-chloride treated sample (\ref{fig:Trajectories}a i-iii) and in the chloride treated sample (\ref{fig:Trajectories}b i-iii). HRTEM movies of both dislocation removals can be found in the SI. The HRTEM images in \ref{fig:Trajectories}a i-iii) show the progression of a $b=\frac{a}{3}\langle2\overline{1}\overline{1}0\rangle$ edge dislocation in non-chloride terminated sample as it is removed. Fourier transform (FT) inlays in each HRTEM image show the spacing and orientation of the $\{1\overline{1}00\}$ family of planes. The presence of non-overlapping frequency spots with 6--fold symmetry in \ref{fig:Trajectories}a i-ii) indicate a mistilt between the $\{1\overline{1}00\}$ planes of the two particles due to the presence of a dislocation at the attachment interface. It can be seen that these sets of non-overlapping spots converge as the dislocation is removed, indicating the formation of a perfect attachment interface. \textit{In situ} HRTEM movies of the imperfect attachment interfaces were analyzed using a custom Matlab routine which determined the location of the dislocation core in each frame (see Methods) revealing the defect trajectories. \ref{fig:Trajectories}a iv) shows the position of the dislocation core as a function of time as it traverses the shortest distance to the surface while exhibiting a mix of glide$/$climb motion. This is consistent with what we previously observed for edge dislocations removed from the interfaces of CdSe NCs attached on via their $\{1\overline{1}00\}$ facets.\cite{Ondry2019}

Next, we examined a chloride-treated sample in order to investigate if the presence of chlorine alters the pathway a dislocation follows to the surface. \ref{fig:Trajectories}b shows the removal of a single $b=\frac{a}{3}\langle2\overline{1}\overline{1}0\rangle$ edge dislocation in the chloride treated sample. HRTEM images (\ref{fig:Trajectories}b i-iii) provide snapshots of the dislocation removal. The extracted dislocation removal trajectory (\ref{fig:Trajectories}b iv) indicates the dislocation in the chloride sample also follows the shortest path to the surface with a similar mixed glide$/$climb character. Comparing these two specific cases, the dislocation present in the chloride terminated sample was removed nearly twice as fast as the dislocation in the non-chloride terminated sample. Importantly, the trajectories of both dislocation cores were similar, indicating the presence of chloride ligands did not affect the dislocation removal \textit{mechanism}, only the dislocation removal \textit{rate}. 

\begin{figure}[htb]
\centering
\includegraphics[scale=1]{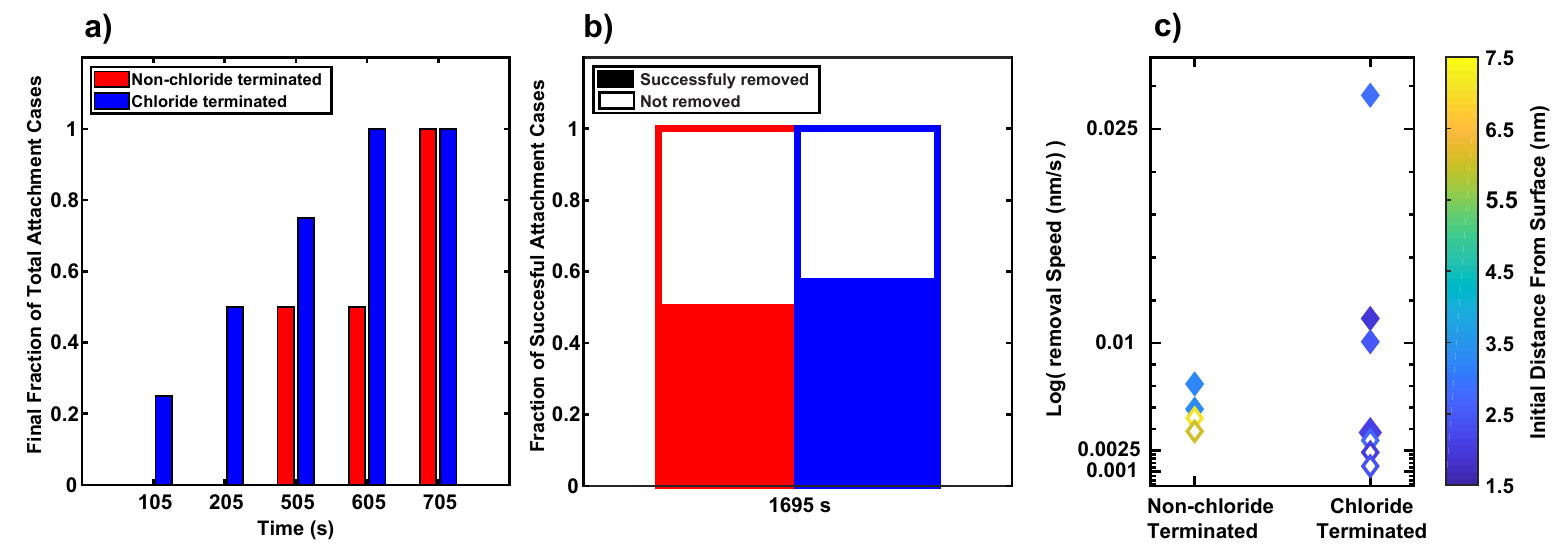}
\caption{ Defect removal times and successful removal fraction along with removal speed distributions for chloride and non-chloride treated samples. a) The number of dislocations successfully removed as a function of time in both the chloride treated (blue) and non-chloride treated (red) samples. b) Final totals of dislocations both successfully removed and not removed for the chloride treated and non-chloride treated samples. c) Dislocation speeds for both successfully removed and not removed dislocations. The filled diamonds represent successfully removed dislocation. The outlined diamonds represent dislocations which were not removed. In this case, the speed was taken as the distance the dislocation core moved in the time the dislocation was observed}
\label{fig:Bar chart}
\end{figure}

To understand the statistical significance of the increased defect removal rate, we analyzed seven HRTEM dislocation removal trajectories for the chloride terminated sample and four HRTEM trajectories for the non-chloride sample. The sample size was limited by the difficulty of finding dimer particle pairs with the correct orientation and crystallographic alignment. \ref{fig:Bar chart}a shows a histogram of the time for successful dislocation removal in both the chloride and non-chloride samples (blue and red respectively). Dislocations were typically removed at shorter times for the chloride treated samples than for the non-chloride treated sample.  However, not all observed dislocations were removed before the sample was destroyed by electron beam irradiation. In these cases, we considered the dislocation removal to be \textit{unsuccessful}. If we consider the ratio of successful to unsuccessful removal cases after 1695s (the longest image series recorded), we observe a nearly identical ratio in both the chloride treated and non-chloride treated samples (\ref{fig:Bar chart}b) with remaining defective interfaces. These results indicate chloride may increase dislocation removal speeds only for those whose local environments allow for complete removal. 

Dislocation removal speeds were calculated for instances where the dislocation was completely removed and where it was only partially removed before sample destruction (filled and unfilled diamond markers respectively). In the former case, the dislocation speed was taken as the distance from the dislocation core to the surface divided by the time it took for complete defect removal. In the latter case, the speed was calculated by measuring the distance the dislocation core traversed at the attachment interface during the time that the NC dimers were imaged. A logarithmic plot of dislocation speeds for each sample as a function of the dislocation core’s initial distance to the surface can be seen in \ref{fig:Bar chart}c.

Comparing the dislocation removal speeds between the two samples shows that the maximum dislocation removal speed observed in the chloride treated sample was nearly twice as fast as the speeds observed in the non-chloride terminated sample. Importantly, these are from identical dislocation types $b=\frac{a}{3}\langle2\overline{1}\overline{1}0\rangle$, whose cores had similar starting distances from the surface ($\sim$4 nm). Additionally, two successful dislocation removal events observed in the chloride terminated sample were found to have faster dislocation removal times than all the non-chloride terminated samples even though both dislocations had larger initial distances from the surface. However, not all dislocation events observed in the chloride treated sample had faster removal times. One successful and three incomplete dislocation removals in the chloride treated sample were found to have either equivalent or slower overall removal speeds compared to the non-chloride treated sample.  This resulted in chloride terminated sample having distribution of removal speeds which was more heavily tailed towards faster speeds that of the non-chloride terminated sample. The approximately equivalent final fractions of successful dislocation removals to unsuccessful removals suggests that chloride surface ligands do not unilaterally make it easier to remove \textit{all} defects. However, the presence of a more heavily tailed final removal speed distribution for the chloride treated sample provides evidence that chloride ligands do alter dislocation removal dynamics and highlights the role local environment plays in successful dislocation removal. 

\subsection{CdSe/CdS Core--Shell 2D NC Array Assembly and Attachment}

Next, we investigated the role of chloride ligands in the remediation of imperfect NC attachment interfaces in 2D superlattices. NC attachment in 2D is of particular interest as it is a crucial step in the synthesis of nanocrystalline materials based on large, ordered assemblies of electronically coupled NC quantum dots.\cite{Sandeep2014,Murray1995} However, the added dimensionality in 2D attachment drastically increases the variety of defects which arise, even in cases where NC size, shape, and orientation can be controlled.\cite{OndryAccount2021} These structural defects can lead to the formation of electronic trap states which prevent the formation of delocalized superlattice electronic bands.\cite{Evers2015,Ondry2018} 

The same CdSe/CdS core-shell NCs used in the dimer attachment studies were employed in the synthesis of these 2D arrays.\cite{Ondry2021} Briefly, the initial NC assembly was synthesized using liquid-air subphase assembly approach\cite{dong_binary_2010} followed by a subphase mediated ligand exchange with thermally labile \textit{t}-BuSH.\cite{Ondry2021} This ligand exchange shortened the interparticle spacing of the assembly.\cite{Ondry2021} The NC assemblies were then transferred onto a TEM grid, allowed to dry, and finally heated to 200$\degree$ C to induce thermal decomposition of the \textit{t}-BuSH ligands, thus defining the unconnected superlattice structure. 

\begin{figure}
\centering
\includegraphics[scale=1]{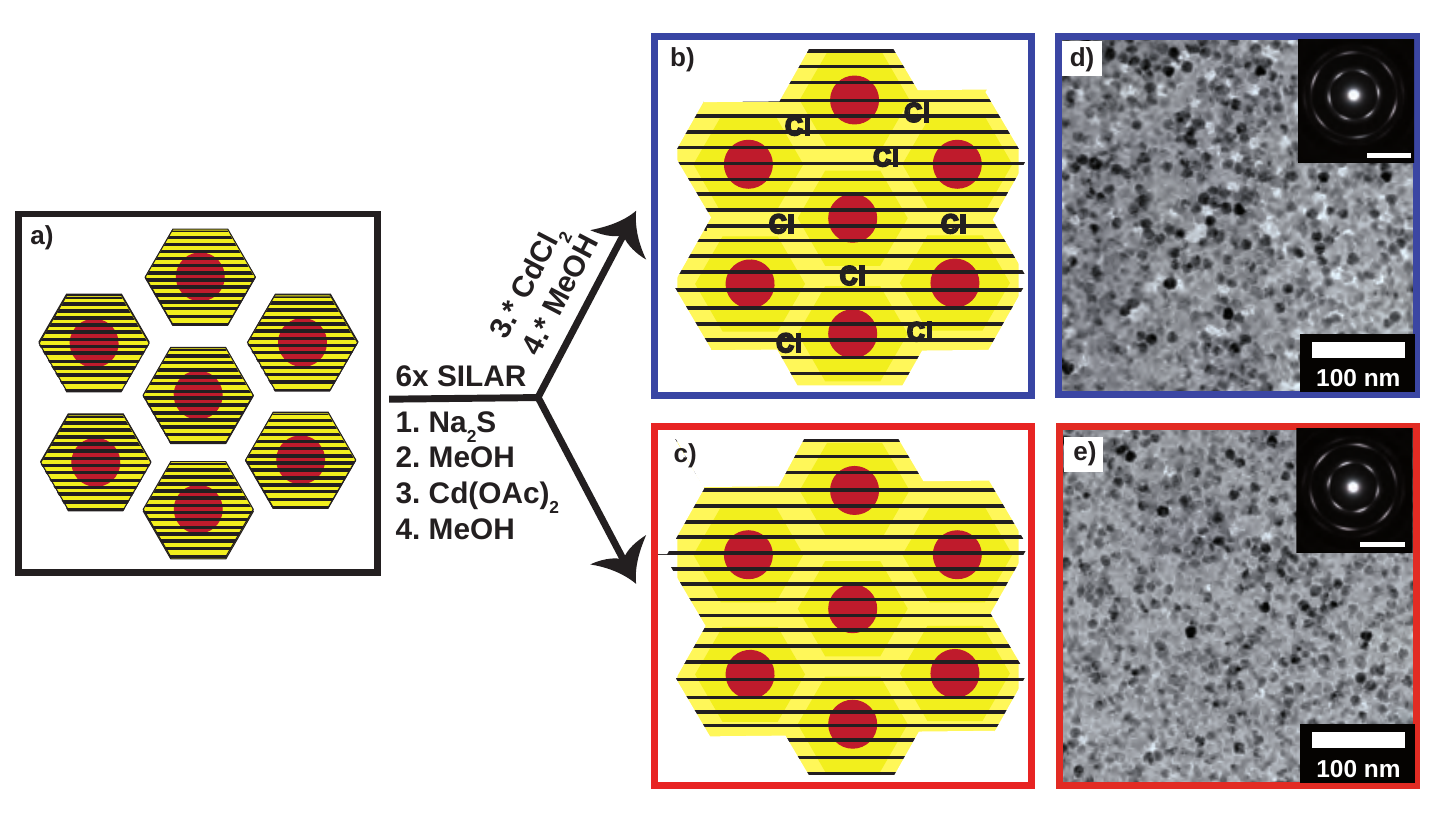}
\caption{ Reaction scheme showing SILAR synthesis of (a) CdSe/CdS core-shell NC assembly. (b) chloride treated epitaxially attached NC arrays and (c) non-chloride treated epitaxially connected NC arrays. (d) chloride-treated NC array post-SILAR and annealing at 300$\degree$C with SAED pattern of a separate HRTEM image with a larger field of view. (e) non-chloride treated NC array post-SILAR and annealing at 300$\degree$C with SAED pattern of larger field of view image. Length of SAED pattern scale bars corresponds to 5 $nm^{-1}$. }
\label{fig:SILAR}
\end{figure}

A reaction scheme showing the synthesis of non-chloride treated and chloride treated NC arrays can be seen in \ref{fig:SILAR}. The NCs were connected by CdS necks using successive ionic layer adsorption and reaction (SILAR) of cadmium and sulfide precursors (see Methods).\cite{Ondry2021} This SILAR method resulted in the simultaneous 2D attachment of the CdSe/CdS NCs. Chloride ligands were introduced to one of the samples by using \ch{CdCl2} as the precursor for the final cadmium layer. It is well established that \ch{CdCl2} passivates electronic defects and promotes grain growth in NC semiconductor thin films\cite{Crisp2014,M.Norman2014}. \ref{fig:SILAR}a shows a diagram of the ordered NC assembly before their attachment via the SILAR method. Following attachment, the NCs individual shell structure was merged via connecting CdS necks (\ref{fig:SILAR}b and c). The arrays were then thermally annealed post-SILAR attachment at 300$\degree$C similar to what was used previously by Ondry et al.\cite{Ondry2021} TEM images showing both the non-chloride treated and chloride treated arrays after thermal annealing are shown in \ref{fig:SILAR}d and e. In both cases, we observe well ordered NC superlattices with single crystal-like electron diffraction patterns. 

To examine the role of chloride surface ions in the 2D epitaxially connected NC arrays, the presence and removal of defects has to be quantified over a large area. As has been previously shown, there exist certain NC interfacial mistilt angles that correspond to the presence of one or more dislocations at the attachment.\cite{Ondry2019} Consequently, the mistilt between the $\{1\overline{1}00\}$ planes of the NCs was taken as an indicator of the presence of defects, such as dislocations, at the attachment interfaces.\cite{Kopp2014} As can be seen in both the HRTEM images and SAED patterns in \ref{fig:SILAR}d and e, the CdSe/CdS core-shell NCs assembled such that their $[0001]$ zone axes were parallel to the incident electron beam. This resulted in the in-plane alignment of their $\{1\overline{1}00\}$ facets. In the case that this alignment had been perfect, the SAED patterns in \ref{fig:SILAR}d and e would show only six spots corresponding to the spacings of the $\{1\overline{1}00\}$ planes. However, six arcs were observed, indicating an imperfect alignment of the NCs along their $\{1\overline{1}00\}$ facets.  

\begin{figure}[htb]
\centering
\includegraphics[scale=1]{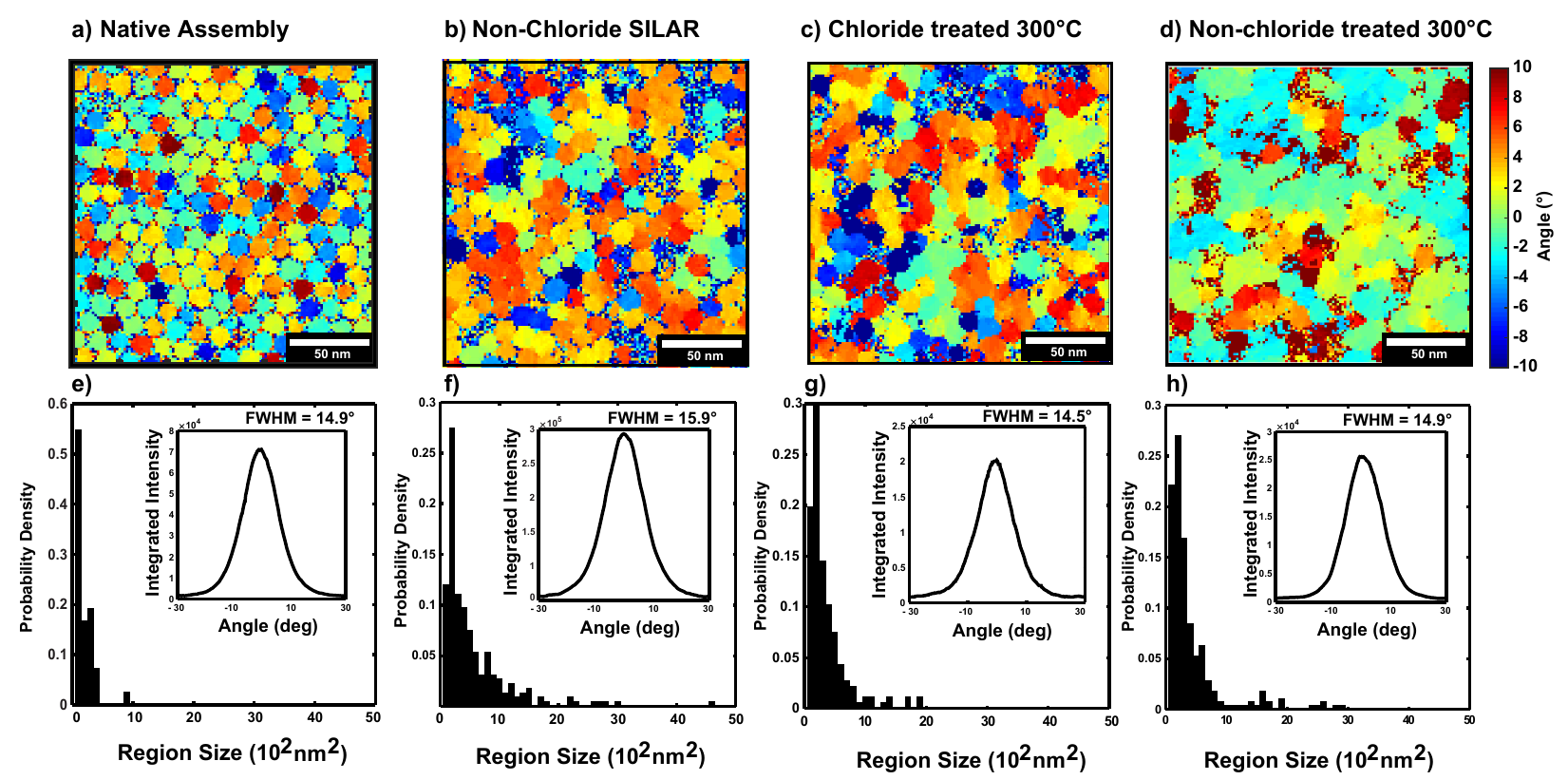}
\caption{Local and global in-plane, NC mistilt as determined by SAED and sliding window Fourier transform analysis. (a) Angle map of 2D NC assembly pre-SILAR. (b) non-chloride epitaxially connected array pre-thermal annealing. (c) Chloride treated epitaxially connected array post-thermal annealing at 300$\degree$C. (d) non-chloride treated epitaxially connected array post-thermal annealing at 300$\degree$C. Common area angle size distribution with an inset of an azimuth angle integrated SAED pattern for native assembly (e), non-chloride treated array post-SILAR pre-thermal annealing (f), chloride treated array post-SILAR and post-thermal annealing (g), and non-chloride treated array post-SILAR and post-thermal annealing (h).}
\label{fig:Angle Maps}
\end{figure}

To better understand the local NC mistilt within areas ($\sim$36,000 $nm^2$) of these arrays, we developed a custom sliding window Fourier transform (SWFT) analysis method to determine the in--plane mistilt topography of the $\{1\overline{1}00\}$ planes from large field of view HRTEM images (See Methods). Representative angle maps from the initial NC assemblies and SILAR fused array pre--thermal annealing, as well as the annealed chloride and non--chloride treated NC arrays can be seen in \ref{fig:Angle Maps}a--d. The angle map of the native NC assembly seen in \ref{fig:Angle Maps}a shows that each unattached NC has a unique $\{1\overline{1}00\}$-plane orientation represented by a solid color. Upon attachment via SILAR, the angle map (\ref{fig:Angle Maps}b) reveals the formation of multi-particle pristine crystallographic regions as indicated by expanded solid color domains. Thermal annealing of both the chloride and non--chloride treated SILAR arrays (\ref{fig:Angle Maps}c-d) shows a loss of individual NC resolution whereby the boundary definition of component particles was reduced with no apparent increase in coherent domain size.

To analyze the angle maps derived from the SWFT code, an additional custom script was designed to determine the relative sizes of regions where all of the NCs were perfectly aligned, eventually obtaining common angle region size probability distributions (\ref{fig:Angle Maps}e--h). A 0.15$\degree$ mistilt threshold was put in place to isolate regions where the NCs’ $\{1\overline{1}00\}$-planes were aligned almost perfectly, yielding a common--angle crystallographic region without dislocations at attachment interfaces. Using the size probability distribution for common angle regions, we were able to more quantitatively compare the native assembly, pre--annealing SILAR, and post--annealing chloride/non-chloride treated SILAR samples. The region size probability distribution for the native assembly (\ref{fig:Angle Maps}e) confirmed our qualitative observation that the native assemblies contained non--attached single NCs of fairly uniform size. SILAR attachment of the NCs resulted in a drastic broadening of the region size probability distribution (\ref{fig:Angle Maps}f), accompanying the formation of multi--particle coherent angle domains. However, the structure of the region size probability distribution narrowed only slightly following thermal annealing in both the chloride and non--chloride treated samples (\ref{fig:Angle Maps}g-h). While SWFT and SAED analysis did not reveal distinct differences between the samples post thermal annealing, raw HRTEM images show noticeable local changes in NC attachment as gaps between the particles are filled and particle borders soften. While HRTEM image analysis provides insight into the local, pristine crystallographic regimes, it does not provide global information about the larger 2D array grains

To understand the mistilt angle statistics over large areas, we collected SAED patterns over an 11 $\mu m^2$ area to obtain global mistilt information about larger NC array domains for all samples. After integrating azimuthally, we averaged the 6 $\{1\overline{1}00\}$ peaks to obtain the \ref{fig:Angle Maps}e--h inlays.  The FWHM of the integrated SAED patterns represented the average mistilt between the $\{1\overline{1}00\}$ planes and is not statistically different between the four samples. This indicates the average mistilt between NCs at the micrometer scale was not improved by SILAR, annealing, or the introduction of chloride ligands which is consistent with our observations from the angle mapping and fourier transform analysis.

In summary, the integrated SAED patterns, angle maps, and common-angle-region size probability distributions allowed us to observe how the superlattice evolved through SILAR attachment, thermal annealing, and the introduction of chloride ligands. SILAR attachment was observed to lock the system in place, preventing rotational mistilt remediation steps. Interestingly, while thermally annealing the arrays post-SILAR was known to increase CdS neck uniformity,\cite{Ondry2021} this step was not observed to yield increased common angle regions nor a narrowed $\{1\overline{1}00\}$ SAED peak (\ref{fig:Angle Maps}f). In contrast to what was observed for 1D attachment, no difference in either the region size probability distribution nor the SAED peak was observed between the chloride and non-chloride treated samples (\ref{fig:Angle Maps}g and h). This suggests that interfacial defect remediation via the introduction of chloride ligands had not occurred in these 2D arrays. More importantly, an analysis of these four systems suggests that the SILAR attachment step prevents NC rotation and the improvement of orientational uniformity. Crucially, this indicates that NC alignment is determined by the assembly step and can not be significantly altered via post-assembly treatment. 

\section{Discussion} \label{Discuss}

\subsection{Understanding the role of chloride ligands increasing dislocation removal kinetics in attached NC dimers} 

The $b=\frac{a}{3}\langle2\overline{1}\overline{1}0\rangle$ Burger’s vector edge dislocations studied in this work had a glide plane offset by 30\degree from the vector normal to the attachment interface.\cite{Ondry2019} Consequently, removal via pure glide motion would have forced the dislocation to travel nearly length--wise through the dimer pair.\cite{Ondry2019} Instead, we observe dislocations in the chloride and non-chloride treated imperfectly attached dimers moving with mixed glide and climb character. The movement of a dislocation \textit{via} partial climb requires atoms to diffuse either towards or away from the dislocation core.\cite{Cai2016} Thus, climb movement is typically slower since it depends on the activation energy barrier for self--diffusion.\cite{Rudolph2017, Anderson2017} In our work, we posit the Lewis acidity of \ch{CdCl2} present on the NCs surface permits \ch{CdCl2} units to form and bind to the surface chalcogenide atoms, as suggested by Owen et al,\cite{M.Norman2014}, enhancing their ability to diffuse. 

\subsection{Understanding the lack of dislocation removal for some interfaces} 

Next, we aim to explain how the local environment can cause some dislocations to be irremovable for both chloride and non-chloride treated samples. It has been our observation that, in some NC dimer attachment cases, at least one of the particles is completely free to rotate (\textit{i.e.,} \textit{only} one attachment). However, particle-to-particle proximity to other particles is an unavoidable component of dropcasting the NCs. As such, there are cases where both dimer NCs are attached to neighboring NCs by one or more facets, though with far greater mobility than found in our 2D NC array systems. As can be seen in \ref{fig:Attachment Scheme}, the presence of an edge dislocation at an interface introduces a mistilt between the two NC lattices. In the case where the two NCs were relatively isolated from or weakly attached to neighboring particles, the NCs had greater freedom to rotate. Thus, dislocation removal alleviates the excess strain energy introduced by the dislocation. In the case where the dimer pair was more interconnected with neighboring NCs, the two NCs had little freedom to rotate during dislocation removal. For this type of NC dimer, the mistilt between the two NCs may be determined by connections to neighboring particles instead of the presence of a dislocation. In these instances, the dislocation may be a result of the enforced mistilt to alleviate strain and thus is not favorable to remove. 

The presence of these two different dislocation removal environments helps explain why not all dislocation removal trajectories observed in the chloride terminated sample had increased speeds and why some dislocations could not be removed even after extensive annealing. In any case, the chloride treatment serves to increase the dislocation removal rate for defects which have a environmentally favored pathway for removal. However for particles which are geometrically frustrated by multiple attachments, the dislocations may be stable due to strain from surrounding particles. 

\subsection{Differences between 1D and 2D NC attachment leading to impaired defect removal for higher dimensionalities}

We found that 2D assembly and SILAR attachment yielded a large increase in common angle region area and thus improved attachment interface quality. However, subsequent annealing and the addition of chloride ligands were not determined to significantly impact attachment interface quality as determined by common angle region size distributions and SAED patterns. This absence of interfacial remediation in the 2D system can best be explained by fundamental differences in NC attachment for 2D and 1D systems. We focus on two key aspects which are significant. The first is the nature of positional and rotational freedom and the second is the proximity of dislocation defects to a free surface. In 1D, there is greater freedom in the positioning of the NCs both pre-- and post--attachment. This freedom can give rise to attachments with larger mistilts, but simultaneously provide the NCs with additional degrees of rotational movement which can facilitate defect removal. In the pre--attachment 1D case, the lack of ordered positioning between neighboring particles can lead to a worse initial alignment between the two NCs.\cite{vanOverbeek2018} However, the same lack of neighboring particles allows the NCs to rotate and align their lattices during dislocation removal.\cite{Ondry2018} Previous work in PbTe NC attachment demonstrated that successful removal of edge dislocations at the NC dimer attachment interface is accompanied by a decrease in the inter-particle lattice mistilt.\cite{Ondry2018} Consequently, the ability of NCs to rotate slightly during defect removal may play an important role in whether a perfect attachment interface can be achieved. 

By contrast, 2D NC arrays possess a significantly lower degree of freedom in NC positioning and rotation by design.\cite{vanOverbeek2018, Ondry2021} The initial assembly of the NCs is a reversible process by which the NCs are all aligned along adjacent facets and have limited variation in their in-plane lattice rotation and position.\cite{vanOverbeek2018} Once equilibrated, the assembly is then locked in place by simultaneous NC SILAR attachment. In this case the NCs are attached to neighboring NCs on all sides and have limited rotational mobility. This likely impairs the system’s ability to remove defects as the rotation which accompanies inter-NC lattice alignment is constrained.

In terms of the proximity of defects to a free surface, the distance the dislocation must traverse to reach a free surface in the 1D case is dictated by the size of the attached NCs and thus is on the scale of a few nanometers. This is important as there are strong elastic forces driving a dislocation to the surface and subsequently facilitating removal.\cite{Ondry2018, Anderson2017, Cai2016} By contrast, the distance between a dislocation core and available surface can vary greatly in 2D NC arrays and depends explicitly on the shape of the NCs and the type of superlattice synthesized. There are certain shapes which tile all space when used as building blocks in an array. Examples of such shapes used in 2D NC arrays are square and hexagonal prisms.\cite{Ondry2021,vanderBurgt2018} In these NC arrays, attachment between NCs occurs on all sides, leaving each individual NC completely connected to the array. As a result, the only free surfaces which exist in the plane of the array are at the edges of the superlattice or at voids in the material. For NCs which have truncated shapes, such as PbSe cuboctahedra used to form honeycomb lattices, the NCs attach only along certain directions.\cite{vanOverbeek2018} As a result, there are free surfaces present within a few nm of the majority of the attachment interfaces.\cite{vanOverbeek2018} Consequently, the absence of available free surfaces in a 2D NC array studied in this work may limit the system’s ability to collectively co-align the NC atomic lattices and attachment post-SILAR treatment. 

\subsection{Possible Pathways to Synthesizing Low Defect Concentration 2D NC Arrays}

In the design of future approaches to achieve highly ordered and defect-free epitaxially connected NC arrays, we recommend a selection of NC building blocks which allow for a high degree of in--plane orientational order to be achieved during assembly. Further we recommend a shape which forms NC superlattices with periodic free surface voids upon oriented attachment. Our work has indicated that it is challenging to correct for interparticle mistilt post--attachment either through thermal annealing or chemical treatments. Consequently, developing NC synthesis and assembly techniques which maximize the degree of in--plane NC alignment will prove a crucial step to synthesizing relatively defect--free 2D NC arrays. Additionally, we propose that designing superlattices which have a built--in proximity between NC attachment interfaces and free surfaces will allow for a more facile removal of dislocation defects.\cite{Ondry2018} This has previously been accomplished by the use of truncated cubic PbSe NCs connected by their \{100\} facets, which yielded a honeycomb hexagonal NC superlattice with a periodic pattern of NCs and free space voids.\cite{vanOverbeek2018}

It should be noted that there is an interplay between the size and shapes of NCs which lend themselves to high order assembly and those that attach along selective facets to yield periodic voids in the superlattice. To synthesize arrays with a high degree of positional and rotational order, it is helpful to use NCs which have interlocking shapes, such as hexagonal prisms.\cite{Ondry2021,ondry_trade-offs_2022} However, as seen in this work, these NCs tile all of space and upon attachment limit the presence of in-plane surfaces to the edges of the array. Conversely, NCs with truncated shapes that allow for void formation do not have this same interlocking between columns and rows, which makes maximizing positional order more challenging. Furthermore, larger NCs lead to greater in-plane orientational order.\cite{ondry_trade-offs_2022}  However, larger particle sizes results in longer diffusion pathways to free surfaces. As a result, it is likely a compromise between multiple synthetic parameters will need to be found in order to synthesize NC arrays which maximize both NC in--plane alignment and have relatively low defect concentrations. 

\section{Conclusion}

In this work, we examined the impact of chloride additives on defect removal dynamics in imperfectly attached NCs with 1D and 2D topology. In the 1D attachment case, \textit{in situ} HRTEM was used to determine the removal speeds and trajectories for edge dislocations present in both the chloride and non--chloride treated samples. Defects in the chloride treated samples were found to have much faster maximum removal speeds than those in non--chloride treated samples. However, this removal rate increase was not universal. Several dislocations in the chloride treated samples were found to have equivalent defect removal speeds to those in the non--chloride treated sample or were not removed at all. The final fraction of successfully removed dislocations to unsuccessfully removed dislocations was nearly identical in both samples. In both the chloride and non-chloride treated samples, the lack of removal in roughly 50\% of observed dislocations is suggested to be the result of differing local environments surrounding the dimers which can prevent defect removal. The increase in removal rates in the chloride-treated samples, when the local environment permits defect removal, is posited to be the result of the \ch{CdCl2} units binding to surface chalcogenide atoms and promoting increased self-diffusion.

Further, we explored chloride additives as a means of facilitating defect removal from 2D arrays of attached NCs where defect concentration is a known barrier to the observation of novel electronic behavior. Attachment via SILAR was observed to induce the formation of regions with perfect crystallographic alignment on the order of 3,000 $nm^2$. We found thermal annealing made no noticeable impact on the size of common angle regions nor the interparticle mistilt as determined by SAED. Likewise, no distinct difference was observed between the chloride and non--chloride treated 2D NC arrays. From this, we suggest that the SILAR attachment step locks the NC arrays into place introducing geometric frustration which prevents any interparticle rotation, preventing the enlargement of common angle regions. This extended geometric frustration is not found in 1D attachment where interface remediation via the introduction of chloride ligands was observed. The drastic contrast between 1D and 2D NC attachment dynamics, highlighted by these experiments, demonstrates the difficulty in removing defects once formed in 2D NC superlattices. To address the loss of NC rotational freedom observed post--SILAR attachment in 2D, we propose that interparticle mistilt \textit{via} minimization in reversible assembly should be prioritized over precise NC shape/ size control. Additionally, introducing built-in free surfaces \textit{via} the use of NCs with truncated shapes may also be critical for facilitating defect removal. We suggest applying these guidelines will help move the field closer to synthesizing NC superlattices with as-predicted emergent electronic phenomena. 

\section{Methods} \label{Methods}

\subsection{Materials}

Cadmium oxide (CdO, 99.99\% Aldrich); octadecylphosphonic acid (ODPA, 99\%, PCI Synthesis); trioctylphosphine oxide (TOPO, 99\%, Aldrich); n-trioctylphosphine (TOP, 97\%, Strem); octadecene (ODE, 90\%, technical grade Aldrich); oleylamine (OAm, technical grade 70\% Aldrich); dioctyl ether (OE, 99\%, Aldrich); oleic acid (OA, 90\%, technical grade Aldrich); cetyltrimethylammonium chloride (CTAC, $>$95\%, TCl); ammonium sulfide (40-48 wt.\%, technical grade Aldrich); anhydrous \ch{Na2S} (Strem); \textit{tert}-butylthiol (\textit{t}-BuSH, Aldrich); cadmium chloride ($>$99.9\%, Aldrich). 

\subsection{Nanocrystal synthesis} 

The CdSe cores and CdS shells were synthesized according to previous methods.\cite{Ondry2021} (See SI for details). 

\subsection{Chloride ligand exchange} 

Chloride ligands were introduced to the CdSe/CdS core--shell NC sample using modified literature procedures.\cite{Saruyama2010} The original core--shell NC sample in hexane (0.6 mL) was loaded into a three-necked flask along with 5 mL dioctyl ether, 400 $\mu L$ OAm, 400 $\mu L$ OA, and 300 $\mu L$ of a 10 mg/mL CTAC/isopropanol solution. The flask was connected to a Schlenk line and degassed for 10 min at room temperature and for an additional 10 min at 50$\degree$C to remove remaining isopropanol. The system was then heated under an argon environment at 200$\degree$C for 1 h. After being cooled to room temperature, the nanocrystals were purified by precipitating with ethanol and centrifugation.The precipitate was dissolved in toluene and used for subsequent experiments.

\subsection{Nanocrystal dimer attachment} 

NC attachment was induced using a modified surfactant ligand removal literature procedure.\cite{Zhang2011} Briefly, a toluene nanocrystal solution was first drop cast onto an amorphous carbon--coated TEM grid and allowed to slowly evaporate. Anti--capillary, self--closing tweezers were then used to pick up the TEM grid and submerge it in a 22 $\mu M$ ammonium sulfide in MeOH solution for 30 s. The TEM grid was then washed via immersion in pure MeOH for 30 s. This attachment procedure was used for both TEM grids containing CTAC--treated nanocrystals and TEM grids containing nanocrystals with their native OA/OAm ligands. 

\subsection{Nanocrystal 2D assembly} 

Nanocrystal assembly was performed at the liquid--air interface following a modified literature procedure.\cite{dong_binary_2010} 1 mL of anhydrous DMF was pipetted into a 1 $cm^3$ Teflon well. 100 $\mu L$ of a dilute CdSe/CdS core--shell nanocrystal solution in octane was carefully pipetted onto the DMF subphase. A glass slide was used to cover the Teflon well and facilitate the slow evaporation of the solvent over a period of 8 hours. 100 $\mu L$ of a solution of 0.17 mM \textit{t}--BuSH in DMF was carefully added to the subphase and allowed to rest for 2 h. Tweezers were then used to insert an amorphous carbon--coated TEM into the Teflon well and scoop upwards, transferring the 2D nanocrystal array onto the TEM grid. The TEM grid samples were dried at 50$\degree$C in a vacuum oven in order to remove any remaining DMF. These dried sample grids were then annealed at 200$\degree$C for 5 min on a hot plate to thermally decompose the \textit{t}-BuSH ligand.

\subsection{SILAR and Annealing} 

The successive ion layer adsorption and reaction procedure followed our previous methods with modifications.\cite{Ondry2021} A 20 mM solution of \ch{Na2S} in MeOH and a 20 mM solution of \ch{CdOAc2} in MeOH were prepared. The ordered 2D nanocrystal array sample prepared via subphase liquid assembly was first submerged in the \ch{Na2S} solution for 30 s and was then washed in pure MeOH for 30 s. Next, the sample was submerged in the \ch{CdOAc2} solution for 30 s and again washed in MEOH for 30 s. To generate the non--chloride treated nanocrystal arrays, this sequence of sample exposure and washing was repeated six times. To generate the chloride treated nanocrystal arrays, the sequence of sample exposure and washing was repeated five times. The sample was then dipped in the \ch{Na2S} solution for 30 s and then in pure MeOH for 30 s. Next, it was submerged in a saturated methanol solution of \ch{CdCl2} for 30 s then washed for a final time in MeOH for another 30 s using a modified literature procedure.\cite{Jain2019} The sample was then heated under vacuum to remove any subphase contaminants. Both the chloride and non--chloride treated SILAR samples underwent thermal annealing at 300$\degree$C for 20 min in a tube furnace under flowing argon. 

\subsection{TEM imaging, Electron Diffraction, and X-Ray Diffraction} 

A FEI Tecnai T20 S--TWIN TEM equipped with a \ch{LaB6} filament was used to obtain \textit{in situ} HRTEM movies, HRTEM images, and electron diffraction patterns for the imperfectly attached CdSe/CdS core--shell nanocrystal dimers and larger assembled arrays. The TEM was operated at 200kV and was coupled to a Gatan Rio 16IS camera with 4k--by--4k resolution. 

\subsubsection{\textit{In situ} HRTEM movies}

For the \textit{in situ} HRTEM movies, images were collected at a rate of 4 fps with a nominal 400 kx magnification. This yielded a final pixel size of 0.017 nm. A custom Gatan Micrograph program was used to control the electron fluence of the electron beam incident upon the sample.\cite{Hauwiller2018} For all but one of the movies, the edge dislocation removal studies were performed at an electron fluence of 8500 $e^-/\AA^2s$. The chloride--terminated movie used in the dislocation trajectory analysis was collected at 8500 $e^-/\AA^2s$.

\subsubsection{HRTEM 2D array images}

For the imaging of the 2D SILAR samples, HRTEM images were taken at a nominal magnification of 145 kx. This correlated to a pixel size of 0.046 nm. All images were taken near Scherzer focus.

\subsubsection{X-ray photoelectron spectroscopy}

X-ray photoelectron spectroscopy (XPS) was performed on a Thermo Fisher Scientific K-Alpha Plus XPS instrument. Survey X-ray photoelectron spectra were obtained by running 10 scans in the constant analyzer energy (CAE) mode with an energy resolution of 1.0 eV. High resolution scans were run for the Cl 2p, N 1s, and Cd 3d peaks using 50 scans in the CAE mode, yielding a final energy a resolution of 0.1 eV. The peaks in the XPS spectrum were fit using XPSPeak 4.1.\cite{PhysRevB.64.205420} The Cl 2p peak was identified by comparing its effective binding energy to known literature values.\cite{NIST} The concentration of chloride ions was determined using $C = I/J \sigma \zeta T \lambda = I/JF $ where the atomic sensitivity factor (F) accounted for all spectrometer and specimen specific terms. The atomic percent chloride was then found using $ a\% = (I_{Cl}/F_{Cl})/(\sum_{n} I_{n}/F_{n}) $. The sample was treated as a four--component system consisting of Cd, Se, S, and Cl. This analysis confirmed the presence of chloride in the sample with a final atomic percent concentration $0.34\%$.

\subsection{Dislocation trajectory determination} 

To obtain dislocation trajectories, the $b=\frac{a}{3}<2\overline{1}\overline{1}0>$ HRTEM movies were drift corrected using the ''\textit{in-situ} player'' in Gatan Micrograph (which uses a cross correlation method) and then converted to an image stack. Each frame of the HRTEM image stack was then cropped to contain only two imperfectly attached NC. The NC were both oriented on the $\langle0001\rangle$ zone axis and a $b=\frac{a}{3}<2\overline{1}\overline{1}0>$ edge dislocation was found at the interface between their $\{1\overline{1}00\}$ facets. We used custom Matlab script to identify the dislocation position in each frame. Briefly we calculated the fast Fourier transformation (FFT) to the cropped image. A circular, binary mask was applied to the FFT to isolate the frequency points containing the spacing information of the set of $\{1\overline{1}00\}$ planes in which the dislocation was found. The inverse FFT of this masked FFT generated a fourier--filtered real space image of the imperfectly attached nanocrystals which only featured the planes containing the edge dislocation. The image was then binarized and skeletonized which preserved the connectivity of the initial image while removing background noise. After this image processing, the edge dislocation presented as a line which abruptly terminated as either a branch or end-point. This termination point was taken as the position of the dislocation core and was tracked through each frame of the HRTEM movies, yielding a map of dislocation’s trajectory.

\subsection{Angle Map Generation}

To obtain an angle map, HRTEM images of the nanocrystal samples both pre-- and post--SILAR treatment were processed using a sliding--window Fourier transformation. A custom sliding--window Fourier transform script was implemented using Matlab. This was accomplished by extracting a 2n by 2n subarea of the image (n = 64 pixels or 2.96 nm), multiplying it by a 2D hanning window, and then calculating its FFT. For each FFT, the high intensity peaks which corresponded to the $\{1\overline{1}00\}$ plane spacings were identified using a Fast Peak Find algorithm.\cite{Natan} The centroid of the peak was determined with subpixel accuracy using a Taylor expansion of a 3 by 3 window around the maximum intensity pixels and determining where the derivative would be zero. The angle each of these intensity peaks made with respect to the horizontal x--axis was determined. Next, the average angle from 2 of the $\{1\overline{1}00\}$ spots was calculated, taking into account the 60$\degree$ rotational symmetry of the lattice. The minimum, positive angle was taken as the rotational orientation of a nanocrystals' $\{1\overline{1}00\}$ lattice planes. This process was repeated for the whole image to generate a 2D rotation map of the $\{1100\}$ lattice plane angles. 

\subsection{2D Sub-grain size analysis} 

In the SILAR--treated samples, the angle maps generated by the sliding-window Fourier transform allowed for the categorization of crystalline regions with a common lattice orientation. 2D sub-grain size analysis was performed to determine the sizes of these common-angle regions and their size-probability distributions in the chloride and non-chloride treated samples. The angle maps were binned according to their measured mistilt--value. A recursive function was introduced which looked at the mistilt values of the pixels surrounding the bin edges to prevent imposing artificial bin sizes. A pixel was included and the bin adjusted if it was determined to belong to a common-angle region. A common--angle region was defined as a region in which difference in nearest neighbor pixel mistilt--angles was less than 0.15\degree. The regions within each bin were converted to image objects and their size and location were determined. The final size of these regions in nanometers was determined from the magnification of the initial HRTEM image.

\section*{Author Information}

\subsection*{Notes}

The authors declare no competing financial interest

\begin{acknowledgement}

This work was supported by the National Science Foundation, Division of Materials Research (DMR), under Award Number DMR-1808151. J.C.O gratefully acknowledges the support of the Kavli Philomathia Graduate Student Fellowship. J. J. C. and M.F.C gratefully acknowledges the National Science Foundation Graduate Research Fellowship under Grant DGE 1752814. We acknowledge Yi-Hsien Lu for assistance in doing the XPS measurements shown here. 
\end{acknowledgement}

\begin{suppinfo}

The Supporting Information is available free upon request from authors. 

\end{suppinfo}

\bibliography{ref.bib}

\end{document}